\documentclass{iau}

\usepackage{graphicx}
\usepackage{natbib}
\usepackage{epsfig,epsf}
\usepackage{array}
\usepackage[english]{babel}

\def\apj{\textit{ApJ}}                 
\def\apjl{\textit{ApJ}}                
\def\apjs{\textit{ApJS}}               
\def\aap{\textit{A\&A}}                
\def\aaps{\textit{A\&AS}}              
\def\mnras{\textit{MNRAS}}             
\def\actaa{\textit{AcA}}             
\def\pasp{\textit{PASP}}               
\def\pasj{\textit{PASJ}}               
\def\memsai{\textit{MemSAIt}}

\def\jaavso{\textit{JAAVSO}}

\newcommand{\Msolar}{\,$M_\odot$}        
\newcommand{\Rsolar}{\,$R_\odot$}        
\newcommand{\Lsolar}{\,$L_\odot$}        

\title[Pulsations in extreme helium stars]{The origin and pulsations\\ of extreme helium stars\footnote{This is the original and complete 
version of a paper based on a review talk given at IAU Symposium No. 301, {\it Precision Asteroseismology}. 
To meet page limits, the version to be published in the Symposium 
Proceedings omits sections 2 and 3.}}
\author[C. S. Jeffery]
       {C. Simon Jeffery} 
\affiliation{Armagh Observatory, College Hill, Armagh BT61 9DG, Northern Ireland, UK\\
}

\pubyear{2014}
\volume{301}  
\pagerange{1--8}
\jname{Precision Asteroseismology}
\editors{J.A. Guzik, W.J. Chaplin, G. Handler \& A. Pigulski, eds.}
\begin{document}
\maketitle

\begin{abstract}
Stars consume hydrogen in their interiors but, generally speaking, their surfaces continue to
contain some 70\% hydrogen (by mass) throughout their lives. Nevertheless, many types of star can be
found with hydrogen-deficient surfaces, in some cases with as little as one hydrogen atom in 10\,000.
Amongst these, the luminous B- and A-type extreme helium stars are genuinely rare;  only
$\sim15$ are known within a very substantial volume of the Galaxy.

Evidence from surface composition suggests a connection to the cooler R\,CrB variables
and some of the hotter helium-rich subdwarf O stars. Arguments currently favour an origin in the
merger of two white dwarfs; thus there are also connections with AM\,CVn variables and Type Ia
supernovae. Pulsations in many extreme helium stars provide an opportune window into their interiors. These
pulsations have unusual properties, some being ``strange'' modes, and others being driven
by $Z$-bump opacities. They have the potential to deliver distance-independent masses and to provide a
unique view of pulsation physics. 

We review the evolutionary origin and pulsations of these stars, 
and introduce recent progress and continuing challenges.

\keywords{stars: early-type, stars: chemically-peculiar, stars: supergiants, stars: white dwarfs, 
stars: evolution, stars: oscillations, stars: variable: other, stars: individual: V652\,Her, FQ\,Aqr}  
\end{abstract}

\section{Extreme helium stars}

Observing from McDonald Observatory at a maximum altitude of 14$^\circ$, \citet{popper42} reported
the B2 star HD\,124448 to ``show no hydrogen lines, either in absorption or in emission, although the
helium lines are sharp and strong ... The abundance of hydrogen appears to be very low in the
atmosphere of this star''. Indeed, HD\,124448 turned out to be the first of around 20 B- and early A-
supergiants, with apparent magnitude $9.3 < V < 12.6$, in which hydrogen comprises less than 1 part
per thousand of the atmosphere. In terms of kinematics, metallicity and galactic distribution, they
have the properties of the Galactic bulge \citep{jeffery87}. 
With luminosities approximately ten thousand times solar,  the number
count is complete for the observable parts of the Galaxy (fainter stars would lie beyond). In
addition to their low hydrogen abundances, the true ``extreme helium stars'' (EHes) show atmospheres
which are enriched in nitrogen by a factor ten, carbon by between one and three parts per hundred,
and sometimes oxygen by a similar amount. The combination of extreme surface composition 
and extreme
rarity makes these stars interesting, and presents a challenge for the theory of stellar
evolution.

\begin{figure}[!ht]
\begin{center}
 \includegraphics[width=12cm]{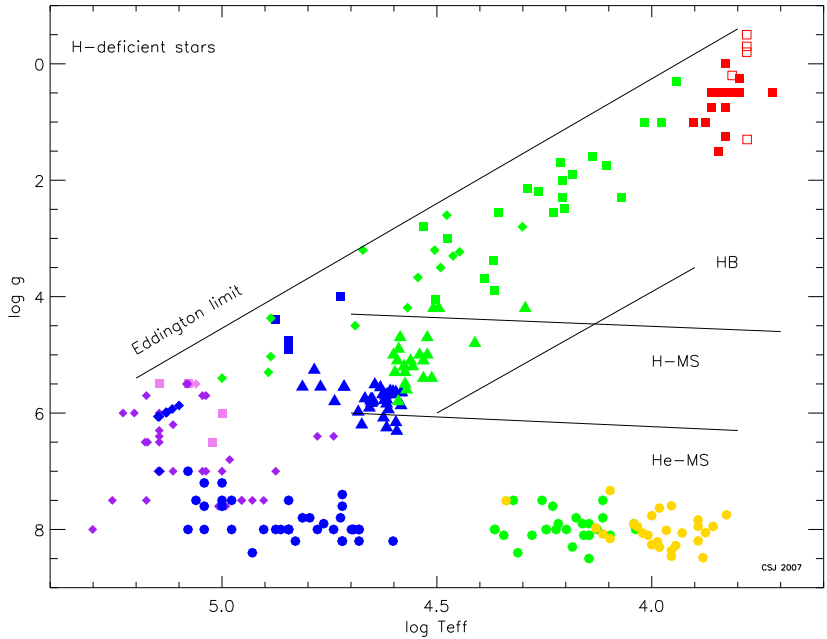} 
\caption{$\log g-\log T_{\rm eff}$ diagram for several classes of low-mass
  hydrogen-deficient stars described in detail by \citet{jeffery08.hdef3.a}. EHe stars are denoted by squares (light grey, green online),
as are RCB (dark grey, red online), HdC (open squares), HesdO$^+$ (back, blue online) and O(He) (mid-grey, violet online) stars.}
\label{f:hdef}
\end{center}
\end{figure}

The general properties of EHes have prompted suggestions of a connection to the cooler R\,Coronae
Borealis stars \citep{schoenberner75}, which are better known for their spectacular and unpredictable light
variability \citep{pigott97}. Figure \ref{f:hdef} shows the distribution by surface gravity and
effective temperature of several classes of hydrogen-deficient star; stars falling on an imaginary
line parallel to the Eddington limit would have the same luminosity-to-mass ratio, corresponding to
the evolutionary path of a giant contracting to become a white dwarf. Links between non-variable
hydrogen-deficient carbon (HdC), R\,Coronae Borealis (RCB), extreme helium (EHe), luminous
helium-rich subdwarf O (HesdO$^+$) and O(He) stars have all been suggested at one time or another.

In order to  establish the  origin of EHes and related objects, temperatures and gravities are required. 
Additional observables are provided by the surface composition and by the fact that many 
EHes are pulsating. Pulsations provide an opportunity to make direct measurements of mass and radius.  
This paper  reviews the spectroscopic data on surface composition, summarizes the principal 
theories for the evolutionary origin, and discusses the major pulsation properties. It concludes with a synopsis of the 
extraordinary pulsating EHe V652\,Herculis. 

\section{Asteroarch{\ae}ology}

The depleted hydrogen surfaces of EHes indicate that a mixture of nuclear-processed helium and carbon 
has been exposed. Abundance analyses have been carried out on the basis of optical and ultraviolet
spectroscopy using primarily LTE methods
 \citep{hill65,schoenberner74,heber83,jeffery92,jeffery93b,drilling98,pandey01,pandey06a,behara06}; 
some partial non-LTE analyses have been executed  \citep{jeffery98,przybilla05,pandey11}.
The majority of these results can be summarised  as follows \citep{jeffery11a}:
\begin{itemize}
\item[--]H  is a relic of the original star and only appears as a trace element.
\item[--]Ca, Ti, Cr, Mn and Ni scale with Fe and are essentially unprocessed.
\item[--][N/Fe] is proportional  to [(C$^\prime$+N$^\prime$+O$^\prime$)/Fe] where primes refer to a primordial value, 
indicating that the original carbon and oxygen has been converted to nitrogen via the hydrogen-burning CNO-cycle.
\item[--][C/Fe]\,$\gg0$, indicating the presence of 3$\alpha$ product.
\item[--][O/Fe]\,$\gg0$, indicating $^{12}$C+$\alpha$ and/or $^{14}$N+$\alpha$ product.
\item[--][Ne/Fe]\,$\gg0$, indicating $^{14}$N+$2\alpha$ product.
\item[--]Mg, Si, S, ... are unremarkable and scale with Fe.
\item[--][F/Fe]\,$\gg0$ \citep{pandey06c}; fluorine is a by-product of the $^{14}$N+$\alpha$ reaction. 
\item[--]Y, Zr (s-process) overabundances are measured in a few cases \citep{pandey04}, 
              indicating material from a thermal-pulsing phase on the Asymptotic Giant Branch (AGB).
\item[--][P/Fe]\,$\gg0$; possible phosphorus overabundances are not understood, but may also have an AGB origin.
\end{itemize}

Together, these  results point to the co-existence of material produced by CNO-cycled hydrogen burning, and 
by 3$\alpha$,   $^{12}$C+$\alpha$ and $^{14}$N+$\alpha$ helium burning. There is  evidence from  RCBs 
that the large oxygen excesses are due to a short-lived phase 
of $^{14}$N+$\alpha\rightarrow^{18}$O burning \citep{clayton07}. 
There are also relics of primordial hydrogen and, 
in some stars, s-process elements  from a previous AGB phase. 
A theory for the origin of EHes must be able to 
explain the  co-existence of all of these products on their surfaces.

\section{Origin}

\citet{schoenberner86} reviewed extant proposals for the origin of RCBs and EHes. 
Several authors have investigated the evolution of single helium stars from the helium main-sequence 
to become helium-shell 
burning giants \citep{paczynski71,biermann71,dinger72,roeser75,weiss87a}. 
Such models mimic the general properties of 
RCBs and EHes, but generally fail 
to explain the detailed surface mixtures -- even if it could be established how the helium main-sequence
stars would form in the first place. 

\citet{paczynski71} had proposed a model involving ``deep envelope mixing'' on
the asymptotic giant branch, which fell short of establishing a mechanism to drive the necessary mixing. 
\citet{scalo75} introduced the notion of ``hot bottom burning'' in which a vigorous nuclear-burning shell 
is sufficiently luminous to drive convection to the stellar surface. Meanwhile, \citet{wood73} and \citet{schoenberner77}
concluded that EHes ``can be understood as the remnants of red giant stars that have lost their hydrogen-rich
envelope during their ascent along the asymptotic giant branch.'' All three proposals struggle to  
fully account for the observed surface chemistries. 

\subsection{Late and very late thermal pulses}
\citet{iben83} proposed that stars which experience a late thermal pulse (LTP, or helium-shell flash)
after leaving the AGB would return to the cool giant regime and there be identified as RCB stars.
There would be mixing between the hydrogen-rich surface layers, the helium-rich intershell, and
helium shell-flash products which would become visible at a now hydrogen-deficient surface.
The principles behind the LTP had been previously predicted
by \citet{fujimoto77a} and demonstrated by \citet{schoenberner79}. In fact, two cases had been
identified, which were later described more fully by \citet{herwig99}. In a {\it late} thermal pulse,
generally on the constant-luminosity part of the post-AGB contraction curve, there is no mixing
between H-rich and He-rich matter (and hence no production of s-process elements) at the time of the
flash; helium is dredged up as the star cools and surface convection penetrates into the helium-rich
layers from above. In a {\it very late} thermal pulse, generally when the star is on or almost on the
white dwarf cooling track, the helium-shell flash is more violent and flash-driven convection mixes
helium- and carbon-rich material directly to the surface. Both models have some attraction for the origin of
RCB stars but have more weaknesses than strengths \citep{clayton11}. They are important in
discussing the origin of ``born-again'' stars like FG\,Sge (late thermal pulse, \citet{jeffery06}),
V4334\,Sgr (very-late thermal pulse \citet{herwig01}), and hydrogen-deficient PN central stars  and
PG1159 stars \citep{werner06}.

\subsection{Double white-dwarf mergers}
At around the same time as the LTP model emerged, \citet{webbink84} and \citet{iben84} realised that a binary system
containing two white dwarfs with a period less than about two hours should lose orbital energy by the
radiation of gravitational waves \citep{landau58}. 
Subsequently the white dwarfs would collide in a timescale less than
the Hubble time. In the case of small mass ratios, the collision will likely be in the form of stable
Roche-lobe overflow, identified  as a possible origin for the  AM\,CVn  stars, which are low mass
hydrogen-deficient white-dwarf binary systems \citep{nelemans01}. 
With a mass ratio closer to unity, the rate of mass loss would be such that the
rate of radius increase of the mass-losing white dwarf ($R_{\rm WD}\propto M_{\rm WD}^{-1/3}$)  leads to
run-away mass transfer, the destruction of the less massive white dwarf, and its immediate
accretion onto the surface of its more massive companion. The overall outcome depends on the composition 
of the white dwarfs and the total mass of the merged product. For example,
a helium white dwarf (He-WD) accreted onto a carbon-oxygen white dwarf (CO-WD) with a total 
mass less than the Chandasekhar mass results in the ignition of
shell-helium burning at the helium/carbon boundary, and the expansion of the star to become a
supergiant, with a hydrogen-poor surface, as in the RCB stars. A more massive merger 
could produce a Type Ia supernova. This connection beween RCB stars and SN Ia, already
 recognised by \citet{wheeler78}, makes the study of double white-dwarf mergers,  
and of RCB and EHe stars,  of considerable interest  \citep{dan13}. 

Since the predictions by \citet{webbink84} and \citet{iben84},  there have been several 
strands to calculations of white-dwarf mergers. Briefly, these include:
\begin{itemize}
\item[i)] Hydrodynamic Calculations. Early simulations
demonstrated that the donor white dwarf would disrupt on a dynamical timescale and rapidly form a
hot corona and a Keplerian disk surrounding the accretor
\citep{benz90,segretain97,guerrero04,yoon07,loren09,longland11,shen12,staff12,menon13,dan13}.
Only recently have large grids of simulations been completed, and only recently have simulations included a detailed
calculation of nucleosynthesis during the merger. An important question is whether these mergers
are ``hot'' or ``cold''; i.e. is the initial helium-shell ignition sufficiently hot and of sufficient duration
to drive nucleosynthesis  (e.g. $^{14}{\rm N}(\alpha,)^{18}{\rm O}$) that would explain the
more exotic aspects of EHe and RCB chemistry. 
A significant result is that, because the donor has further to fall into the potential well, mergers with 
large mass ratios ($q\approx 1$) are hotter and therefore more likely to produce $^{14}$N+$\alpha$ 
products than mergers with smaller mass ratios.
\item[ii)] Post-Merger Evolution. \citet{saio02} showed that a 0.6 or 0.7{\Msolar} CO-WD
accreting between 0.1 and 0.3{\Msolar} of helium at half the Eddington rate would ignite said
helium and expand to produce a giant with the mass, luminosity, temperature and basic composition of
an RCB star. After shell helium exhaustion, these models contracted with properties corresponding to
observations of EHe stars. The detailed surface chemistries could be explained in an {\it ad hoc}
fashion using a simple recipe based on the pre-merger composition of the various hydrogen-rich,
helium-rich and carbon-rich layers of the progenitor white dwarfs. With a decade of additional
abundance data for both EHe and RCB stars, \citet{jeffery11a} extended this simple recipe; in
particular they suggested that the fluorine, neon and phosphorus overabundances could originate in
the AGB intershell of the CO-WD, with $^{18}{\rm O}$ coming from a pocket close to the carbon-helium
boundary in the CO-WD (in their models). This, together with near-normal abundances of magnesium,
points to progenitor main-sequence stars with initial masses in the range 1.9--3{\Msolar}.
\citet{staff12}, on the other hand, argue that most of the excess exotics can be produced during a
``hot'' merger. 
\item[iii)] Population Statistics. The  expectation of producing sufficient DWD mergers to
populate the Galaxy with RCBs and EHes is obtained by comparing the expected numbers of 
double white dwarfs, with appropriate masses and periods, with the gravitational-radiation
orbital-decay times. The galactic merger rate can then be multiplied by the 
predicted  lifetimes of post-merger stars as RCBs and EHes. The first qualitative 
estimates were made by \citet{iben90}, and confirmed by \citet{nelemans01} to give a CO+He
merger rate of  $\approx 2-5\times10^{-3}$~${\rm yr}^{-1}$. With a contraction (thermal) 
timescale of $\approx 300 - 3000$~yr, there should be $\approx  1 - 15$ EHes in the Galaxy, 
consistent with the observed number. However, \citet{han98} showed that  
binary-star population synthesis (BSPS)  contains  large error margins, so the apparent
agreement should be taken with a pinch of salt. A significant problem still to be resolved is that
the BSPS calculations show most CO+He mergers at the current epoch should come from 
relatively young  systems ($<1$ Gyr), which is inconsistent with the Galactic bulge distribution of
EHe and RCB stars. 
\item[iv)] Dynamical Stability. Mass ratio plays a key role in whether
a double white dwarf binary  merges or  transfers matter slowly when one component
fills its Roche lobe \citep{han99}; the primary factor is the rate of change of semi-major axis 
compared with the expansion rate of the mass-losing star, which gives a critical ratio
$q \geq 0.7 + 0.1 (m_2/M_\odot)$ \citep{han99}. 
The question is closely connected  with the efficiency of  mass transfer and
factors such as synchronicity and nuclear reactivity. It is important, as it strongly affects
the population statistics, but appears far from resolved. 
\item[v)] Angular Momentum. The coalescence of two stars in a short-period orbit raises the
questions of (a) whether the accretor can continue to accrete as it spins up and (b) whether the
total angular momentum of the  product would cause it to break up. Assuming angular momentum
to be conserved within the system, there is an expectation that a contracting RCB star would
spin up to beyond breakup \citep{gourgouliatos06}. However if the surface velocity is close to the
Kepler velocity, angular momentum is transported efficiently from the star to the disk so that
accretion continues as long as matter around the star exists \citep{paczynski90,popham91},
and viscous disks are known to be very efficient transporters of angular momentum 
outward and of mass inward
\citep{lyndenbell74}.
\item[vi)] Progenitors. The question of whether low-mass 
short-period double white-dwarf binaries capable of forming EHe stars are produced by binary-star 
evolution has been emphatically answered by the discovery of significant numbers of such systems
\citep{marsh95b,brown11b,kilic11a,kilic12a}. 
\end{itemize}
Curiously, there is a connection between the DWD merger model and the single-star
evolution models of \citet{dinger72} and others. 
\citet{zhang12a} extended the work of \citet{saio00} on double He-WD mergers
to investigate the formation of helium-rich sdO and sdB stars, which are essentially low-mass helium
main-sequence stars. \citet{zhang12b} found that, in a few cases, HesdO stars formed from the most
massive double He-WD mergers would 
become carbon-rich helium giants during a  He-shell burning phase after core-helium depletion. 
However, the He+He channel provides less than 1\% of the  RCB and EHe stars provided by the CO+He  channel.

\begin{figure}[!t]
\begin{center}
 \includegraphics[width=12cm]{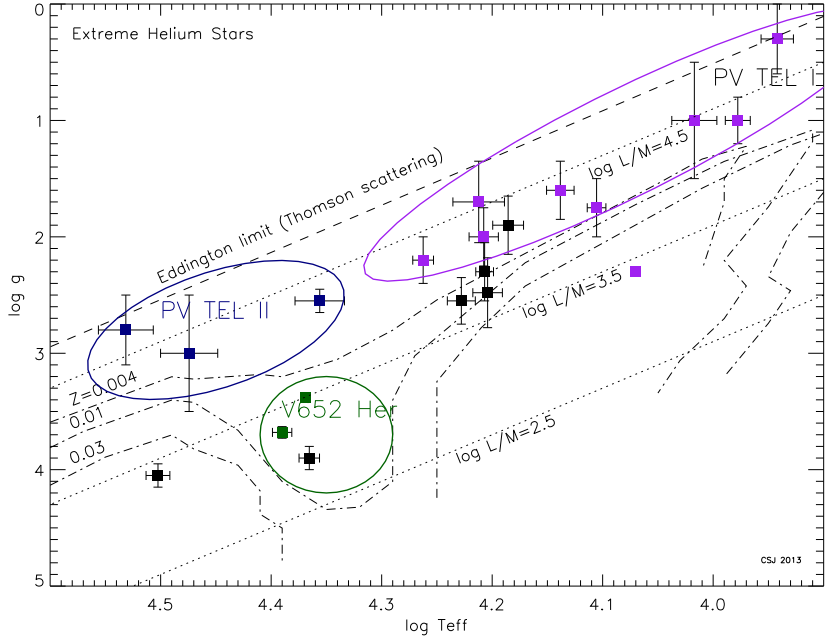} 
 \caption{The $\log g-\log T_{\rm eff}$ diagram for EHe variables, adapted from \citet{jeffery08.ibvs}, 
including the
position of the Eddington limit (assuming Thomson scattering: dashed)
and the loci of stars with given luminosity-to-mass ratios (solar
units: dotted). Stars above the boundaries shown for metallicities
$Z=0.004, 0.01, 0.03$ (dot-dash) are predicted to be unstable to
pulsations \citep{jeffery99a}. 
Ellipses (coloured in electronic version) identify three groups of
pulsating helium stars. In the
electronic version, PV\,TEL\,I variables are shown in purple, PV\,TEL\,II variables in blue, 
and V652\,Her variables in green. Non-variables are black.
 }
   \label{f:puls}
\end{center}
\end{figure}

\section{Pulsation}

\subsection{Discovery and classification}

\citet{landolt75} made the first detection of variability in an EHe, the hot star HD\,160641 = V2076
Oph which showed a brightening by 0.1 mag over seven hours. The discovery of a 0.108\,d pulsation
period in V652\,Her \citep{landolt75} prompted searches for short-period
variability in other EHes. \citet{walker85} reported irregular small-amplitude variations of weeks
to months in HD\,168476 = PV\,Tel, thus codifying the PV\,Tel variables as a class. \citet{bartolini82}
reported a short-period variation in BD$+10^{\circ}2179$ = DN\,Leo which could not be verified
\citep{hill84,grauer84}. From 1983 -- 1988, a St Andrews/SAAO campaign made discoveries of
variability in six EHe stars, namely FQ\,Aqr, NO\,Ser, V2205\,Oph, V2244\,Oph and V1920\,Cyg
\citep{jeffery85a,jeffery86.hdef.b,jeffery85b,morrison87a,morrison87b}. As with DN\,Leo, variability
was not confirmed in Popper's star HD\,124448 = V821\,Cen \citep{jeffery90}. In the following decade,
variability was discovered in V4732\,Sgr, V5541\,Sgr and V354\,Nor \citep{lawson93,lawson98}, and
most recently in the enigmatic MV\,Sgr \citep{percy12}. Following a prediction by \citet{saio95b},
\citet{kilkenny95} discovered 0.1\,d pulsations in BX\,Cir. The distribution of variable and
non-variable EHes is shown in Fig.~\ref{f:puls}. \citet{jeffery08.ibvs} gives a complete list of
properties including original catalogue numbers, GCVS variable star designations, approximate
periods, temperatures, and gravities.

The first discovery of the St Andrews/SAAO campaign concerned FQ\,Aqr, which showed a 21.2\,d
sinusoidal oscillation with an amplitude of 0.4\,mag in $V$ and 0.05\,mag in $b-y$.
\citet{jeffery85a} associated this with pulsation. Together with periods for other EHes, a
pulsation-period effective-temperature relation of the form $\Pi \propto T_{\rm eff}^3$ was
apparent. For stars of the same luminosity, this corresponds to the period mean-density relation for
classical radial pulsators. On the basis of this inference, \citet{saio88b} calculated a series of
linear non-adiabatic pulsation models for low-mass high-luminosity stars in the temperature range
7\,000 to 30\,000\,K. They showed that, for sufficiently high values of $L/M$, opacity-driven
strange-mode radial pulsations would be excited, consistent with the periods observed in the cooler
PV\,Tel variables (Fig.~\ref{f:puls}). \citet{jeffery08.ibvs} calls these ``Type I'' PV\,Tel
variables.

At $T_{\rm eff}>20\,000$\,K, both V2205\,Oph and V2076\,Oph were apparently multiperiodic on
timescales longer than that consistent with a fundamental radial pulsation, and had been inferred to
be g-mode non-radial pulsators \citep{jeffery85b,lynasgray87}. Together with V5541\,Sgr,
\citet{jeffery08.ibvs} calls these ``Type II'' PV\,Tel variables.

The short-period variability in the lower luminosity star V652\,Her was not explained until the
introduction of OPAL opacities \citep{opal92}, when \citet{saio93} showed that, in the absence of
hydrogen, The $Z$-bump opacity mechanism could easily drive radial pulsations with normal (solar-like)
metallicities. \citet{jeffery08.ibvs} puts BX\,Cir and V652\,Her in the same class, but they would
be better labelled V652\,Her variables (Fig.~\ref{f:puls}).

\subsection{Pulsation properties}

What promised to be a class of simply periodic variables with a strict period-temperature relation
and the possibility of measuring direct radii using Baade's method turned out to be a chim{\ae}ra. A
second season of observations of FQ\,Aqr did not reveal a unique period, with 21.5\,d and 23.0\,d
being possible \citep{jeffery86.hdef.b}. \citet{kilkenny99c} monitored both FQ\,Aqr and NO\,Ser for
five consecutive seasons and, from the Fourier power spectrum, reported the ``apparent presence of
several periods but, if real, none seems to persist for more than one season.'' Data from SuperWASP
(unpublished), and a wavelet analysis of the Kilkenny et al.~data are similarly ambiguous. The only
recurring signal occurs at a period of around 20\,d, but is not coherent over a long period of time.

The absence of a regular period is not fatal for extracting stellar radii, so long as the
oscillation is radial and angular radii and radial velocities can be measured simultaneously.
Radial-velocity amplitudes of a few km\,s$^{-1}$ were measured in PV\,Tel, FQ\,Aqr and V2244\,Oph at
the same time as angular radius measurements were obtained with \textit{IUE} \citep{jeffery01a}. Relative
phases were consistent with a radial pulsation. In two cases, the resulting radius and mass
measurements were consistent with theoretical expectation, but the relative mass errors did not
provide a strong constraint for establishing an evolutionary origin.

The aperiodicity of FQ\,Aqr is reflected in observations of other EHe stars \citep[cf.][]{walker85}.
\citet{wright06} considerably extended the photometry and spectroscopy of V2076\,Oph. Instead of
recovering the periods of 0.7 and 1.1\,d reported by \citet{lynasgray87}, they reported that
``conventional Fourier analysis \ldots fails to reveal coherent frequencies'' and suggest that the
light curve ``could be a result of random variations''. On the other hand, the high-resolution
spectrum of V2076\,Oph shows prominent line-profile variability (LPV) on a timescale of a few hours
which is symptomatic of a non-radial pulsation. Of note is that, although LPVs are visible in all
He\,{\sc i} lines, they are only seen in the He\,{\sc ii}\,4686\,{\AA} line, and not in other He\,{\sc ii}
lines \citep[Fig. 2:][]{jeffery08.hdef3.b}. Such intriguing results are almost impossible to
interpret in isolation from photometric monitoring and over such a short time interval.

\begin{figure}[!t]
\begin{center}
\includegraphics[width=12cm]{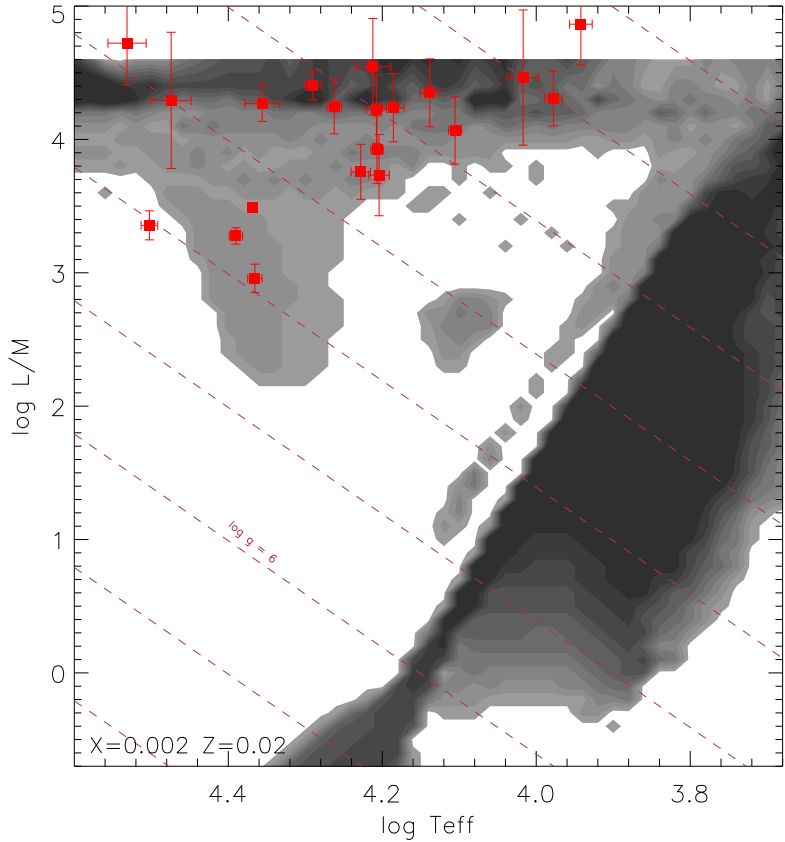} 
 \caption{Contour plot showing the number of unstable radial modes with $n<15$ in linear non-adiabatic pulsation 
analyses of hydrogen-deficient stellar envelopes. The ordinate is $\log L/M$ in solar units, and the abscissa is $\log T_{\rm eff}$. 
The plot is virtually invariant to the mass, at least in the range $0.2 - 1.0$\,{\Msolar}. 
The solid squares  correspond to EHe stars. White means
no unstable modes except for $\log L/M > 4.6$ where model envelopes are difficult to compute. Surface 
gravity contours for $\log g=8,7,6,\ldots$ are represented as broken lines. }
   \label{f:EHe_puls}
\end{center}
\end{figure}

\subsection{Theoretical considerations}

\citet{jeffery13b} extended their investigations of pulsation stability as a function of hydrogen
abundance (mass fraction: $X$) to a larger range of effective temperature ($T_{\rm eff}$),
luminosity-to-mass ratio ($L/M$) and $X$ than before \citep{jeffery99a}. Models for $0.9>X>0.002$
were computed. Figure \ref{f:puls} shows the instability domain for $X=0.002$ and solar metallicity.
At large $X$, the classical Cepheid instability strip is clearly seen. At high $L/M$ strange modes
are excited. As $X$ is reduced, the $Z$-bump instability finger starts to develop, and a strip of
intermediate-order radial modes develops blueward of the classical instability strip. The strange
modes are less sensitive to $X$, and at low $X$ correspond to the observed locations of the variable
EHe stars (Fig.~\ref{f:EHe_puls}). At low metallicity, the $Z$-bump finger is much diminished, and the
lower edge of the strange-mode domain moves to higher $L/M$ ratio (Fig.~\ref{f:puls}), so the
absence of variability in metal-poor EHes is anticipated \citep{saio88b,saio93}. With no hydrogen,
the morphology of the classical strip again changes and other pockets of instability appear, in
particular one at $T_{\rm eff}\approx13\,000$\,K and $L/M\approx500${\Lsolar}/{\Msolar}. It
remains to be seen whether any real hydrogen-deficient stars can be identified with such models.

Strange-mode oscillations in stars with high $L/M$ ratios have been identified for some time
\citep{wood76}. They appear theoretically in non-adiabatic pulsation analyses and have no
corresponding modes in the adiabatic approximation. In particular, they appear to be associated with
stellar envelopes where high opacities in the ionisation zones also lead to a density inversion --
effectively creating a radiation-pressure dominated cavity in the stellar interior. They are
discussed at greater length by, {\it inter alia}, 
\citet{gautschy90} and \citet{saio98b}.

Evidence that non-radial oscillations may be present in the hottest EHes led \citet{guzik06} to find
a large number of unstable opacity-driven g-modes in models for V2076\,Oph. It will be interesting
to learn whether extreme non-adiabacity and strange-mode characteristics or frequent switching
between closely-spaced modes is responsible for the absence of long-lived coherent periods in these
stars.

\section{The shocking case of V652 Herculis}

BD$+13^{\circ}3224=$\,V652\,Her is an extreme helium star and a B-type giant discovered by
\citet{berger63.v652}. With a nitrogen-rich, carbon-poor surface, it pulsates with a period of
0.108\,d which is decreasing at a substantial rate \citep[cf.][]{kilkenny82,kilkenny05}. This period
decrease implies a radius contraction that, together with a radius $2.31\pm0.02$\,{\Rsolar}
and mass $0.59\pm0.18$\,{\Msolar} \citep{jeffery01b}, in turn implies that V652\,Her is
evolving to become a hot subdwarf within $\le10^5$\,yr \citep{jeffery84b}. Single-star models cannot
explain the origin of V652\,Her, which has an almost completely CNO-processed atmosphere
\citep{jeffery99}. On the other hand, models for the post-merger evolution of two helium white
dwarfs very successfully match nearly all of the observational properties \citep{saio00}, as well as
demonstrating that such mergers will become helium-rich hot subdwarfs \citep{zhang12a}.

\citet{saio93} demonstrated that $Z$-bump opacity instability drives fundamental-mode radial
pulsations in V652\,Her. The radial-velocity curve shows that, for nine tenths of the cycle, the
surface layers are nearly ballistic \citep{hill81,lynasgray84,jeffery86}. Close to minimum radius,
the surface acceleration is so large that the atmosphere may be shocked \citep{jeffery01b}. Since
hydrogen-deficient atmospheres are in general more transparent than hydrogen-rich ones, V652\,Her
provides a unique opportunity to study the dynamical behaviour of a pulsating atmosphere at greater
depths (or densities) than is usually the case. Suitable observations could test non-linear
hydrodynamic models for pulsation \citep{fadeyev96,montanes02}. \citet{jeffery13.fuji2} present a
more extended account.

High-speed spectroscopy of V652\,Her was obtained with the Subaru High Dispersion Spectrograph. 2
nights of observations cover over six pulsation cycles with a temporal resolution of 174 seconds.
These observations aim to identify structure in the line cores around minimum radius, to resolve the
passage of the wave through the photosphere using ions with different ionisation potentials, to
determine whether this generates a shock front, or whether the photosphere adjusts subsonically, and
to establish how close the ``free-fall'' phase is to a true ballistic trajectory. The interpretation
of the observations requires coupling a hydrodynamic model of the pulsation to a radiative transfer
code, so that the emergent spectrum can be computed realistically. A summary of progress is given by
\citet{jeffery13.fuji2,jeffery13.sdob6}.

\section{Conclusion}

Extreme helium stars form a group of some 15 low-mass supergiants of spectral types A and B. Their
helium-dominated atmospheres are extremely hydrogen poor ($<0.1\%$) and carbon-rich ($1-3\%$).
Evidence from kinematics, surface composition and distribution in effective temperature and surface
gravity points to a strong link with the cooler R\,CrB stars, and to their origin in the merger of a
helium white dwarf with a carbon-oxygen white dwarf.

Most EHe stars are photometric variables with amplitudes around one tenth of a magnitude. Most are
also radial-velocity variables with amplitudes of a few km\,s$^{-1}$. The timescales of these
variations as a function of effective temperature are consistent with the stellar dynamical
timescales and hence with being due to pulsations. Theoretical models show most EHes to be unstable
to opacity-driven radial pulsations. However, in all but two cases (V652\,Her and BX\,Cir), the
observed variations are {\it not strictly} periodic. Additional work is necessary to better
characterise the periods and amplitudes and to obtain radii using the Baade-Wesselink method.

V652\,Her and BX\,Cir are less luminous than the majority of EHes. V652\,Her can be explained by the
merger of two helium white dwarfs evolving to become a helium-rich subdwarf. The carbon-rich BX\,Cir
is harder to explain, but could be similar. Both stars pulsate radially with periods of 0.1\,d. The
radial-velocity curve of V652\,Her is nearly ballistic, with a very steep acceleration phase. New
observations and models provide a unique dataset and toolkit for exploring the physics of these
pulsations


\end{document}